\documentstyle[12pt]{article}
 \newcommand{\be}{\begin{equation}}
 \newcommand{\ee}{\end{equation}}
 \newcommand{\ba}{\begin{eqnarray}}
 \newcommand{\ea}{\end{eqnarray}}
 \def\entry#1#2{\vbox{\hbox to 92truept{\hrulefill}\break
                \hbox{\vrule\vbox to 30truept{
                \vfill
                \hbox to 92truept{\hfill\quad\small #1\quad\hfill}\break
                \vfill
                \hbox to 92truept{\hfill\quad\small #2\quad\hfill}
                \break\vfill
                \hbox to 92truept{\hrulefill}}\vrule}}}

 \def\arrwv#1#2{\vbox to 68truept{\vfill
                \hbox to 92truept{\put(48,0){\line(0,-1){20}}\hfill}\break
                \vfill
                \hbox to 92truept{\hfill\small #1\hfill}\break
                \vfill
                \hbox to 92truept{\hfill\small #2\hfill}\break
                \hbox to 92truept{%
                \put(48,0){\vector(0,-1){20}}\hfill}\break
                \vfill}}

 \def\arrwh#1#2{\vbox to 30truept{\vfill
                \hbox to 92truept{\small\hfill #1\hfill}\break
                \hbox to 92truept{\rightarrowfill}\break
                \hbox to 92truept{\small\hfill #2\hfill}\break\vfill}}

 \def\d{\partial}

 \def\PRD#1#2#3{{\it Phys.\ Rev.} {\bf D#1}, #2 (#3)}

 \def\NPBP#1#2#3{{\it Nucl.\ Phys.\ (Proc.\ Suppl.)} {\bf B#1}, #2 (#3)}
 
 \def\PLB#1#2#3{{\it Phys.\ Lett.} {\bf B#1}, #2 (#3)}

 \def\IJMPA#1#2#3{{\it Int.\ J.\ Mod.\ Phys.} {\bf A#1}, #2 (#3)}
 \def\IJMPD#1#2#3{{\it Int.\ J.\ Mod.\ Phys.} {\bf D#1}, #2 (#3)}
 \def\MPLA#1#2#3{{\it Mod.\ Phys.\ Lett.} {\bf A#1}, #2 (#3)}

 \def\ANP#1#2#3{{\it Ann.\ Physics (N.Y.)} {\bf #1}, #2 (#3)}

\begin{document}
\title{Geometrodynamical Formulation of Two-Dimensional
Dilaton Gravity and the Quantum Birkhoff Theorem}
\author{\large Marco Cavagli\`a\thanks{Electronic address:
cavaglia@aei-potsdam.mpg.de}\\[3mm]
\em Max-Planck-Institut f\"ur Gravitationsphysik\\
\em Albert-Einstein-Institut\\
\em Schlaatzweg 1, D-14473 Potsdam, Germany}
\date{}
\maketitle
\maketitle
\begin{abstract}
General two-dimensional pure dilaton-gravity can be discussed in a unitary
way by introducing suitable field redefinitions. The new fields are
directly related to the original spacetime geometry and in the canonical
picture they generalize the well-known geometrodynamical variables used in
the discussion of the vacuum dilatonic black hole. So the model can be
quantized using the techniques developed for the latter case. The
resulting quantum theory coincides with the quantum theory obtained
imposing the Birkhoff theorem at the classical level. 
\end{abstract}
\noindent
Two-dimensional pure dilaton-gravity (DG) is described by the action 
\be
S=\int_\Sigma d^2x\sqrt{-g}[U(\phi)R(g)+V(\phi)+
W(\phi)(\nabla\phi)^2]\,,\label{action-gen} 
\ee
where $U$, $V$, and $W$ are arbitrary functions of the dilaton $\phi$, and
$R$ is the two-dimensional Ricci scalar. The model based upon Eq.\
(\ref{action-gen}) has been extensively investigated both from the
classical and quantum points of view because of its connection to string
theory, reduced models, black holes, and membrane physics. (For short
reviews and references see e.g.\ \cite{Filippov,JSM,Cproc} and the
contribution of V.\ de Alfaro in this volume.) Classically one may
choose $U(\phi)=\phi$ and locally set $W(\phi)=0$ by a Weyl-rescaling of
the metric. (Throughout the paper we shall make this choice for
simplicity.) So different choices of $V(\phi)$ identify different
theories. Remarkable examples are the CGHS model \cite{CGHS} ($V =
const$), the Jackiw-Teitelboim model ($V = \phi$), and the
four-dimensional spherically-symmetric Einstein gravity
($V\propto \phi^{-1/2}$).

The model described by Eq.\ (\ref{action-gen}) is completely classically
integrable for any choice of the potential $V(\phi)$. Integrability has
been discussed by V.\ de Alfaro in his talk so here I shall not
enter details. Let me just recall that both the metric and the dilaton can
be expressed in terms of a free field and one invariant parameter which is
a local integral of motion independent of the coordinates \cite{Filippov}. 
This property does not depend from the choice of the dilatonic potential. 
Classical solutions can be written in a closed form without specifying
$V(\phi)$. 

Although the classical properties of the model based upon Eq.\
(\ref{action-gen}) are well-known, not much is known about its
quantization: only the CGHS model and the vacuum dilatonic black hole have
been investigated in depth. The two most fruitful attempts to construct
the quantum theory of dilaton-gravity are described in Refs.\
\cite{Jackiw,CDF} and Ref.\ \cite{Varadarajan} respectively. 

The first approach, used for the CGHS model, is based on a canonical
transformation mapping the original system into a system described by free
fields. Then the theory is quantized in the free field representation. The
main problem of this approach is that the new canonical variables are not
directly related to the original spacetime geometry and important physical
quantities, as the mass, cannot be expressed in terms of the new fields
\cite{CDF}. Further, it is not clear how to generalize the canonical
transformation for an arbitrary dilatonic potential. 

The ``geometrodynamical approach'' was originally developed by Kuch\v{a}r
for the canonical description of the Schwarzschild black hole
\cite{Kuchar}. This approach uses variables that are related to the
spacetime geometry and does not make use of the field redefinitions of
Refs. \cite{Jackiw,CDF}. Only the vacuum dilatonic black hole has been
quantized using this formalism. Indeed, it is not clear how to formulate a
geometrodynamical approach without choosing the functional form of the
dilatonic potential. On the other hand, the classical properties of the
model described by Eq.\ (\ref{action-gen}) do not depend on the choice of
$V(\phi)$ and a single description, irrespective of the potential chosen,
does indeed exists. So at this stage a main question arises: can we
describe by a unique formalism, i.e.\ without specifying the potential
$V(\phi)$, the quantum version of the model based upon Eq.\
(\ref{action-gen})? 

In this short contribution we answer positively to this question and
derive an explicit canonical transformation to geometrodynamical variables
for any choice of the dilatonic potential. This is performed both at the
Lagrangian and canonical levels. Once this has been done, the quantization
is straightforward. The theory is reduced to quantum mechanics and wave
functions coincide with those that are found imposing first the Birkhoff
theorem and then the quantization algorithm as V.\ de Alfaro has outlined
in a previous section of this conference. (See also Refs.\ \cite{bh} for
the case $V\propto\phi^{-1/2}$).

In Ref.\cite{Filippov} Filippov finds a B\"acklund-like transformation
mapping the dilaton and the metric into a free field and a local conserved
quantity
\be
M=N(\phi)-\nabla_\rho\phi\nabla^\rho\phi\,,
\label{transf-M}
\ee
where $N(\phi)=\int^\phi d\phi'V(\phi')$. Here we show
that the action can be written 
\be
S=\int_\Sigma d^2x\,\sqrt{-g}\,{\nabla_\mu\phi\nabla^\mu M
\over N(\phi)-M}+S_{\d\Sigma}\,,\label{action-new}
\ee
where $S_{\d\Sigma}$ is a surface term and $\nabla$ represents the covariant
derivative w.r.t.\ the metric $g_{\mu\nu}$. The key of the proof is the
observation that in two-dimensions $R$ can be written
\be
{R\over 2}=\nabla_\mu\left({\nabla^\mu\nabla^\nu\chi\nabla_\nu\chi-
\nabla_\nu\nabla^\nu\chi\nabla^\mu\chi\over
\nabla_\rho\chi\nabla^\rho\chi}\right)\,,\label{ricci-div}
\ee
where $\chi$ is an arbitrary scalar field. (Equation (\ref{ricci-div}) follows
from the commutator of covariant derivatives and from the definition of
the two-dimensional Riemann tensor after some algebraic manipulation.)
Differentiating Eq.\ (\ref{transf-M}), and choosing $\chi=\phi$, both
$V(\phi)$ and $R$ can be written as functions of $M$ and $\nabla_\mu\phi$.
Finally, by an integration per parts we find Eq.\ (\ref{action-new}).

Equation (\ref{action-new}) has the same number of d.o.f.\ of the original
action. This can be checked using the ADM metric parametrization
$ds^2=\rho[(\alpha dx_0)^2-(\beta dx_0-dx_1)^2]$, where $\alpha$ and
$\beta$ play the role of the lapse function and of the shift vector
respectively. Field equations derived from Eq.\ (\ref{action-new}) are
equivalent to those obtained from Eq.\ (\ref{action-gen})
\ba
&(\nabla_{(\mu}\nabla_{\nu)}-g_{\mu\nu}\nabla_\sigma\nabla^\sigma)
\phi+{1\over 2}g_{\mu\nu}V(\phi)=0\,,\label{eq-dil}\\ 
&R+{dV\over d\phi}=0\,.\label{eq-R} 
\ea 
(Note that Eq.\ (\ref{eq-R}) is automatically satisfied if Eq.\
(\ref{eq-dil}) is satisfied provided that
$\nabla_\rho\phi\nabla^\rho\phi\not=0$. By requiring that the fields
and their derivatives are continuous -- see later -- Eq.\ (\ref{eq-R})
implies Eq.\ (\ref{eq-dil}) everywhere.) Varying the action
(\ref{action-new}) we obtain
\ba
&&\nabla_\mu\nabla^\mu\phi-V(\phi)=0\,,\label{eq-M}\\
&&\nabla_{(\mu}\phi\nabla_{\nu)}M-{1\over
2}g_{\mu\nu}\nabla_\sigma\phi\nabla^\sigma M=0\,,\label{eq-g}\\
&&\nabla_\mu M\nabla^\mu
M+\nabla_\nu\phi\nabla^\nu\phi\nabla_\mu\nabla^\mu M=0\,.\label{eq-phi}
\ea
Equation (\ref{eq-M}) corresponds to the trace of Eq.\ (\ref{eq-dil}). 
Moreover, differentiating Eq.\ (\ref{transf-M}) it follows that Eqs.\
(\ref{eq-g}-\ref{eq-phi}) are satisfied if Eq.\ (\ref{eq-dil}) is
satisfied because Eq.\ (\ref{eq-dil}) implies $\nabla_\mu M=0$.  The
converse latter statement is also true provided that
$\nabla_\rho\phi\nabla^\rho\phi\not=0$.  Further, when this condition
holds Eq.\ (\ref{eq-g}) implies $\nabla_\mu M=0$. So by requiring the
continuity of the fields one obtains that Eqs.\ (\ref{eq-M}-\ref{eq-phi}) 
and (\ref{eq-dil}-\ref{eq-R}) are equivalent. This statement can also be
directly checked using the ADM metric parametrization. 
Solving the field equations in the gauge $\alpha=1$, $\beta=0$ and using
the null coordinates $u=(x_0+x_1)/2$, $v=(x_0-x_1)/2$ we find
\be
M=constant\,,~~~~~~\rho=[N(\phi)-M]\d_u\psi\d_v\psi\,,\label{sol}
\ee
where $\psi$ is a harmonic function related to $\phi$ by the equation
$d\psi/d\phi=[N(\phi)-M]^{-1}$. Equation (\ref{sol}) coincides with the
solution derived from Eq.\ (\ref{action-gen}) \cite{Filippov}.  Equation
$\nabla_\rho\phi\nabla^\rho\phi=0$ identifies the horizon(s)  of the
metric. So the request of continuity introduced below Eqs.\
(\ref{eq-M}-\ref{eq-phi}) amounts to require the continuity of fields
across the horizon(s). 

The canonical formalism is an essential step to quantize the model. 
Starting from Eq.\ (\ref{action-new}) and using the ADM metric
parametrization the action can be cast in the Hamiltonian form.
The super-Hamiltonian and the super-momentum are
\be
{\cal
H}=[N(\bar\phi)-M]\pi_{\bar\phi}\pi_M+[N(\bar\phi)-M]^{-1}\bar\phi'M'\,,~~~~
{\cal P}=-\bar\phi'\pi_{\bar\phi}-M'\pi_M\,,
\ee
where $\pi_{\bar\phi}$ and $\pi_M$ are the conjugate momenta to
$\bar\phi\equiv\phi$ and $M$ respectively and primes represent derivatives
w.r.t.\ $x_1$. Eventually, the canonical action must be complemented by a
boundary term at the spatial infinities. This can be done along the lines
of Refs. \cite{Kuchar,Varadarajan}. The resulting boundary term is of the
form $S_{b}=\sum_{b}\int dx_0 M_b\alpha_b$, where $M_b\equiv
M(x_0,x_1=boundary)$ and $\alpha_b(x_0)$ parametrize the action at
boundaries.  The canonical chart $(\bar\phi,\pi_{\bar\phi},M,\pi_M)$ can
be related to the original canonical variables
$(\phi,\pi_{\phi},\rho,\pi_\rho)$: 
\ba
&M=N(\phi)-\displaystyle{\rho^2\pi_\rho^2-\phi'^2\over\rho}\,,
&\pi_M=\displaystyle{\rho^2\pi_\rho\over 
\rho^2\pi_\rho^2-\phi'^2}\,,\label{tr-M}\\
&\bar\phi=\phi\,,~~~~~~~~~~~~~~~~~~~
&\pi_{\bar\phi}=\pi_\phi-\displaystyle{\rho^2\pi_\rho\over
\rho^2\pi_\rho^2-\phi'^2}\left[V(\phi)+2\pi_\rho\left({\phi'\over
\rho\pi_\rho}\right)'\right]\,.\label{tr-phi}
\ea
After some tedious calculations one can check that the only non-vanishing
Poisson brackets at equal time $x_0$ are
\be
[M(x_0,x_1),\pi_M(x_0,x_1')]=\delta(x_1-x_1')\,,~~~~~
[\bar\phi(x_0,x_1),\pi_{\bar\phi}(x_0,x_1')]=\delta(x_1-x_1')\,,
\ee
so Eqs.\ (\ref{tr-M}-\ref{tr-phi}) define a canonical map.

The canonical field equations and the constraints can be easily solved. 
${\cal H}=0$ and ${\cal P}=0$ are solved by $\pi_{\bar\phi}=0$, $M'=0$. 
Clearly, this solution has the same physical content of Eqs.\
(\ref{sol}). The second Eq.\ (\ref{sol}) is a direct consequence of the
canonical field equations. 

Using the new canonical chart the Dirac quantization of the model is
straightforward.  $\pi_{\bar\phi}=0$ implies that the wave function do not
depend on $\bar\phi$. Finally, in the $M$ representation the
eigenfunction of $M(x_1)$ with eigenvalue $m$ is
$\Psi(M(x_1);x_0)=\chi(m,x_0)\delta(M(x_1)-m)$, where $\chi(m,x_0)$
satisfies the Schr\"{o}dinger equation with reduced Hamiltonian given by
the boundary action. This completes the quantization of the model. 

Let us conclude with few remarks. We have derived a canonical
transformation to geometrodynamical variables that generalizes the
transformation of Ref.\ \cite{Varadarajan} to any DG model.  The general
DG action can be cast into the form (\ref{action-new})  and the system can
be described both in the canonical and Lagrangian framework using the new
variables. This is a new and interesting result that may open the way to
sundry applications. Thermodynamics and functional quantization of the
model starting from Eq.\ (\ref{action-new}) are two remarkable examples.
The quantization of the general DG model becomes straightforward in the
new variables and the resulting Hilbert space coincides with the Hilbert
space obtained imposing the Birkhoff theorem at the classical level.
Quantum field theory reduces to quantum mechanics and the model exhibits
at the quantum level the Birkhoff theorem. 
\section*{Acknowledgements}
I am grateful to the organizers of the {\it XI Conference on Problems of
Quantum Field Theory} for hospitality and financial support. I am indebted
to Vittorio de Alfaro and Alexandre T.\ Filippov for many interesting
discussions about the subject of this paper.  This work has been supported
by a Human Capital and Mobility grant of the European Union, contract no.\
ERBFMRX-CT96-0012.

\end{document}